\documentclass[11pt,a4paper]{article}

\usepackage[margin=1in]{geometry}
\usepackage[T1]{fontenc}
\usepackage[utf8]{inputenc}
\usepackage{graphicx}
\usepackage{amsmath}
\usepackage{subcaption}
\usepackage{upgreek}
\usepackage[version=4]{mhchem}
\usepackage{microtype}
\usepackage{placeins}
\usepackage{booktabs}
\usepackage{siunitx}
\usepackage{array}
\usepackage{caption}
\usepackage{xcolor}
\usepackage{float}
\usepackage{csquotes}
\usepackage{fancyhdr}
\usepackage[colorlinks=true, linkcolor=blue, citecolor=blue, urlcolor=blue]{hyperref}

\newcolumntype{L}[1]{>{\raggedright\arraybackslash}p{#1}}
\providecommand{\linenumbers}{}
\providecommand{\nolinenumbers}{}
\newcommand{\sficsfigtwoWidth}{0.82\linewidth}
\renewenvironment{abstract}
  {\begin{center}\bfseries Abstract\end{center}\begin{quote}\small}
  {\end{quote}\vspace{1em}}
\makeatletter
\patchcmd{\thebibliography}{\section*{\refname}}{\section*{\refname}\vspace{-0.6em}}{}{}
\apptocmd{\thebibliography}{%
  \small
  \setlength{\itemsep}{0.2em}%
  \setlength{\parskip}{0pt}%
}{}{}
\makeatother
\AtBeginDocument{\colorlet{red}{black}}

\begin{document}

\begin{center}
{\LARGE\bfseries A Spatial-Resolved Proton Energy Spectrometer Based on a Scintillation-Fiber Cube\par}
\vspace{0.8em}
{\normalsize
Tan Song$^{1,2}$, Ying Gao$^{1,2,*}$, Di Wang$^{3}$, Yujia Zhang$^{1}$, Jiarui Zhao$^{1,2}$, Qingfan Wu$^{1}$, Zhuo Pan$^{1}$, Shirui Xu$^{1}$, Ziyang Peng$^{1}$, Yulan Liang$^{1}$, Tianqi Xu$^{1}$, Zihao Zhang$^{1}$, Haoran Chen$^{1}$, Qihang Han$^{1}$, Xuan Liu$^{1}$, Ye Yang$^{3}$, Maocheng Wang$^{3}$, Siguang Wang$^{1}$, Yihua Yan$^{3}$, Zhongming Wang$^{3}$, Wenjun Ma$^{1,2,4,\dagger}$\par}
\vspace{0.6em}
{\small\itshape
$^{1}$State Key Laboratory of Nuclear Physics and Technology, School of Physics, Peking University, Beijing 100871, China\par
$^{2}$Beijing Laser Acceleration Innovation Center, Beijing 101407, China\par
$^{3}$National Key Laboratory of Intense Pulsed Radiation Simulation and Effect, Northwest Institute of Nuclear Technology, Xi'an 710024, China\par
$^{4}$Institute of Guangdong Laser Plasma Technology, Baiyun, Guangzhou 510540, China\par}
\vspace{0.4em}
{\small $^{*}$Corresponding author: \href{mailto:ying_gao@pku.edu.cn}{ying\_gao@pku.edu.cn}\quad
$^{\dagger}$Corresponding author: \href{mailto:wenjun.ma@pku.edu.cn}{wenjun.ma@pku.edu.cn}\par}
\end{center}
\vspace{0.8em}

\begin{abstract}
Advanced particle acceleration methods have produced high-peak-current ion beams with broad energy spread and complex spatial distribution. There is an urgent need to develop online spatial-resolved energy spectrometers for high-energy pulsed ions. This paper introduces a novel spectrometer based on a scintillation-fiber-cube for the purpose of online diagnosing proton beams with broadband energy spread and complex spatial distribution. We present its working principles, experimental setup, and comprehensive calibration using monoenergetic and spatially uniform proton beams generated by a synchrotron accelerator. \textcolor{red}{Calibration results confirm an energy measurement range of 6--93 MeV with relative energy uncertainty of 0.6\% at 80 MeV, and a pixel size of 0.5 mm for beam profile reconstruction.} By exploiting a custom-designed energy degrader, we generated a complex proton beam and measured it with the scintillation-fiber cube spectrometer (SFICS). The results demonstrated the spectrometer's potential for online measurement of the energy spectrum and spatial distribution of complex proton beams.
\end{abstract}

\noindent\textbf{Keywords:} Laser-driven proton acceleration; proton spectrometer; scintillation fiber; online detector

	\section{Introduction}\label{sec:intro}
    \nolinenumbers	
{\color{red}
\begin{table*}[htbp]
	\centering
	\small                          
	\setlength{\tabcolsep}{4pt}     
	\renewcommand{\arraystretch}{1.2}
	\caption{\textcolor{red}{Comparison of representative scintillator-based proton spectrometers with spatial resolution capability.}}
	\label{tab:spectrometer_comparison}
	\color{red}
	\begin{tabular}{@{}%
			L{0.19\textwidth}
			L{0.15\textwidth}
			L{0.12\textwidth}
			L{0.18\textwidth}
			L{0.15\textwidth}
			L{0.14\textwidth}
			@{}}
		\toprule
		\textbf{Spectrometer} & \textbf{Spatial Type} & \textbf{Energy Channels} & \textbf{Energy Range} & \textbf{Pixel Size} & \textbf{Energy Uncertainty} \\
		\midrule
		Metzkes et al. (2016) \cite{Metzkes2016} & 2D & 9 channels & \SIrange{1}{20.1}{\mega\electronvolt} & $4 \times 4~\si{\square\milli\meter}$ & Not specified \\
		Marine et al. (2024) \cite{Huault2024}  & 2D & 10 channels & \SIrange{4.1}{20}{\mega\electronvolt} & $93 \times 200~\si{\square\micro\meter}$ & Not specified \\
		Metzkes et al. (2012) \cite{Metzkes2012} & 1D & 9 channels & \SIrange{1.6}{18.3}{\mega\electronvolt} & \SI{1.3}{\milli\meter} & Not specified \\
		SFICS (this work)     & Quasi-2D & 120 channels & \SIrange{6}{93}{\mega\electronvolt} & $0.5 \times 0.5~\si{\square\milli\meter}$ & 0.6\% at \SI{80}{\mega\electronvolt} \\
		\bottomrule
	\end{tabular}
\end{table*}
}
\linenumbers
\nolinenumbers
\begin{figure*}[!t]
	\centering
	\includegraphics[width=1\hsize]{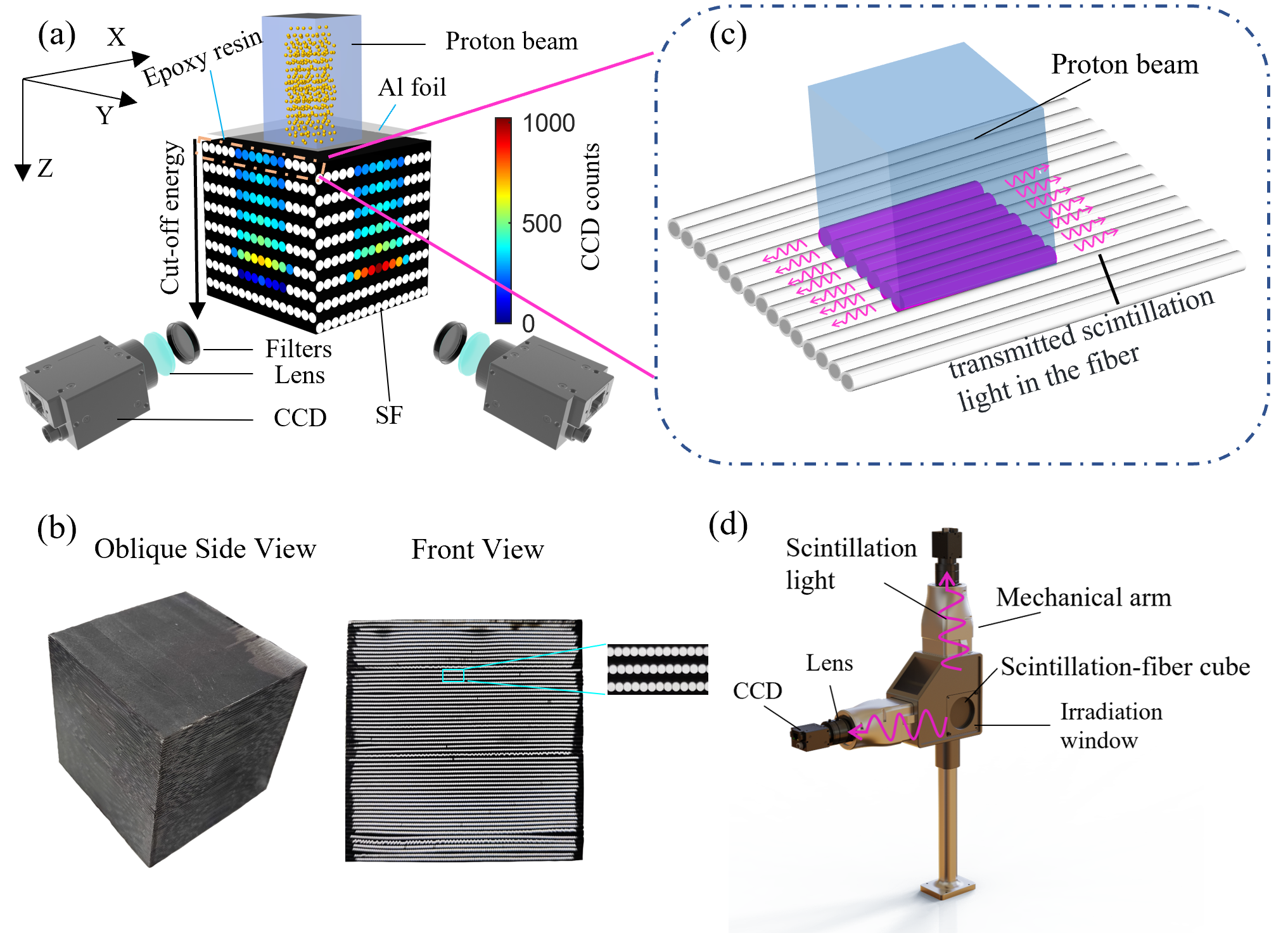}
	\caption{(a) Schematic of the structure and working principle of the scintillation-fiber-cube spectrometer. (b)Photos of the scintillation-fiber-cube: oblique side view, front view and enlarged detail of frontal view.  (c) Schematic illustration of the production and transmission of scintillation photons in a layer of scintillation fibers caused by a proton beam. (d) Mechanical assemble of the SFICS.}
	\label{fig:1}
\end{figure*}	
\linenumbers	
	High-energy proton beams have a wide range of applications in various fields, including proton therapy\cite{Mohan2017, Mohan2022, Yang2023}, isotope production\cite{Schmor2011, Griswold2016, Jiang2024}, and fundamental nuclear physics research\cite{Matsushita2016, Liu2024, Jiang2025}. To ensure optimal performance in these applications, various diagnostic devices, such as intensity monitors\cite{Unser1981, Ghergherehchi2010, Giordanengo2022, Zhang2022, Donegani2023, Zhao2017}, energy spectrometers\cite{Leeper1986, Nesteruk2019, Levin2024}, and profile monitors\cite{Bosser1985, Blumer1995, Hori2008, Levasseur2017, Xing2018}, have been developed to characterize the beam properties. They are suited for continuous beams generated by radio frequency quadrupole or cyclotron accelerators, which typically are monoenergetic with regular spatial distribution. Recently, ion acceleration via laser plasma interaction has achieved significant progress. Protons with maximum energies up to 150 MeV have been generated\cite{Ziegler2024}, indicating its promising potential for medical applications. Unlike conventional proton beams, laser-driven proton beams are bunched in time and non-uniform in space. \textcolor{red}{Moreover, their energy spectra are typically exponential and angularly chirped\cite{Schollmeier2014, Dromey2016, Ma2021}.} These characters impose huge challenges for the beam diagnostic methods and devices. \textcolor{red}{Some detectors have been used in experiments to characterize laser-driven protons including time-of-flight spectrometers (TOF)\cite{Margarone2008, Milluzzo2017, Milluzzo2019, Scuderi2020, Ngai2024}, Thomson parabola spectrometers (TPS)\cite{Prasad2010, Jung2011, Prasad2013, Gwynne2014}, radiochromic films (RCF) stacks\cite{Schollmeier2014, Nurnberg2009, Kaufman2015, Xu2019}, Colum-bia Resin \#39 (CR-39) nuclear-track detectors\cite{Zhang2019, He2020, Zhang2020} and radioactivation stacks\cite{Gunther2013, Shou2023}.} TOF and TPS can measure the energy spectra instantly, but they are unable to provide spatial information of the beams due to their small acceptance angles. Radioactivation stacks, CR-39 detectors, and RCF stacks are passive detectors capable of withstanding extreme environmental conditions such as intense wideband electromagnetic pulses (EMPs), which generally enhances their performance. These detectors provide excellent two-dimensional spatial resolution. Specifically, RCF offers high resolution for high-energy protons due to its thin structure; CR-39 provides single-particle sensitivity and exceptional spatial resolution; while radioactivation stacks feature a wide dynamic range and immunity to electronic noise. However, their inability to provide real-time diagnostics significantly reduces experimental efficiency.
		
	With the emergence of high-repetition-rate ultra-intense lasers, there has been an increasing need for online spatial-resolved proton energy spectrometers\cite{Metzkes2012, Metzkes2016, Dover2017, Huault2019, Schwind2019, Mariscal2021, Hesse2022, Huault2024}. Scintillator detectors are well-suited for online diagnostics of laser-driven proton beams due to their stability, excellent radiation resistance\cite{Xu2012}, and large dynamic range.\textcolor{red}{Several types of scintillator-based spatially resolved energy spectrometers have been reported; a comprehensive comparison of representative designs is presented in Table \ref{tab:spectrometer_comparison}.} Metzkes et al.\cite{Metzkes2016} developed a pixel energy spectrometer by employing a differential-filtering-masked scintillator to create macro-pixels, achieving 9 energy channels and \textcolor{red}{a pixel size of $4 \times 4~\si{\square\milli\meter}$}. \textcolor{red}{The spectrometer achieves good energy uncertainty performance but with limited spatial sampling granularity.} Marine et al.\cite{Huault2024} reported a novel scintillator stack spectrometer which arranges scintillators in an accordion-like configuration, allowing each scintillator to be imaged from the side. \textcolor{red}{The achieved pixel size was $93 \times 200~\si{\square\micro\meter}$.} Due to the large gaps (~2.4 cm) between the scintillator layers, the size of the whole spectrometer is inevitably large, which is detrimental to experiments. Moreover, the distortion of the beam due to the scattering on the scintillator layers is amplified in the gaps. Therefore, the number of scintillator layers is restricted. Metzkes et al.\cite{Metzkes2012} presented a spectrometer based on tightly packed plastic scintillator plates with transverse signal readout. It can provide 1D-resolved energy spectra, but \textcolor{red}{the spatial sampling is limited to $\SI{1.3}{\milli\meter}$ pixel size}.

	\textcolor{red}{In this paper, we report a proton spectrometer based on a scintillation-fiber cube that simultaneously achieves low energy uncertainty and fine spatial sampling. This spectrometer, designated as \enquote{Quasi-2D} in Table \ref{tab:spectrometer_comparison}, can reconstruct 2D spatial-energy distributions under specific conditions, such as weak spatial-energy correlation or with auxiliary collimation, rather than providing direct full 2D detection.} The scintillation-fiber cube consists of 120 orthogonally packed scintillation fiber layers. Each layer is composed of 120 seamlessly packed scintillation fibers. Along the beam incident direction, the SFICS provides 120 channels for energy spectra measurements. Meanwhile, it provides the opportunity to reconstruct the beam's spatial characteristics by utilizing transverse scintillation images. We calibrated the SFICS using monoenergetic and spatially uniform proton beams generated by a synchrotron accelerator. Thereafter, we applied the SFICS to measure an artificial complex proton beam. The results preliminarily demonstrate its potential for online measurement of the energy spectrum and spatial distribution of complex proton beams.

	\section{Structure and Working Principle of the SFICS}\label{sec:principle}
	This section describes the structural design and working principles of the Scintillation-Fiber-Cube Spectrometer (SFICS). By integrating a custom-made scintillation-fiber cube with a pair of orthogonally arranged imaging systems, the SFICS can provide spatially resolved profile diagnostics and energy deposition information of the incident proton beam, enabling comprehensive characterization of proton beam energy spectra, particularly in transient radiation environments.
	\subsection{Structure}\label{subsec:structure}

Figure~\ref{fig:1} provides a comprehensive overview of the SFICS. As shown in Figure~\ref{fig:1}(a), the SFICS consists of two subsystems: a scintillation-fiber cube and a pair of orthogonally arranged imaging systems. An aluminum foil of appropriate thickness is positioned at the beam entrance to block ambient stray laser light and unwanted heavy ions produced during laser plasma acceleration. \textcolor{red}{The scintillation-fiber cube (photos shown in Fig.~\ref{fig:1}b) has overall dimensions of $60~\mathrm{mm} \times 60~\mathrm{mm} \times 60.5~\mathrm{mm}$ and is fabricated as a cubic block by casting 120 orthogonally stacked layers of plastic scintillation fibers (each $0.5~\mathrm{mm}$ in diameter) in black epoxy resin. This fiber diameter defines the pixel size for spatial beam profile measurements.} 
This layered orthogonal design enables independent measurements of 1D transverse profiles of the beam in two orthogonal directions (X and Y), providing spatial distribution information of deposited energy projected along the X and Y axes, respectively. As shown in Fig.~\ref{fig:1}c, the number of scintillation photons output by each scintillation fiber is determined by the total radiation energy deposited in that scintillation fiber.

\textcolor{red}{The black epoxy resin not only serves as the casting matrix but also plays a crucial role in minimizing optical crosstalk between fibers and enhancing spatial resolution. Monte Carlo simulations confirm that optical crosstalk between adjacent fibers is less than 0.1\%.} Each layer comprises 120 0.5-mm-diameter plastic scintillation fibers. They have the advantages of high radiation hardness, low cost, and low density. The relatively low stopping power of the plastic fibers is beneficial for high-energy resolution. The scintillation efficiency of the fibers is 8000 photons/MeV. Their characteristic emission at 435 nm has a good spectral match for common charge-coupled device (CCD) detectors.

Each imaging system consists of a lens, a CCD camera, and several optical filters positioned in front of the CCD to record the scintillation signals transmitted from the XOZ and YOZ planes of the cube. The optical filters are placed according to the experimental needs.

\textcolor{red}{The CCD cameras feature 10-bit grayscale depth with 5.5~$\mu$m pixel size, and a maximum frame rate of 30 fps. Scintillation signals are quantified as \enquote{CCD counts}, representing averaged grayscale values within each fiber's region of interest.}

Figure~\ref{fig:1}d shows the mechanical assembly rendering which connects the scintillation-fiber cube and the imaging system. The arms are light-tight by a shell made of aluminum alloy. If the ambient radiation is too strong, additional lead sheets can be wrapped outside the shell to protect the CCD.

    \nolinenumbers	
	\begin{figure}[!htb]
		\centering
		\includegraphics[width=\sficsfigtwoWidth]{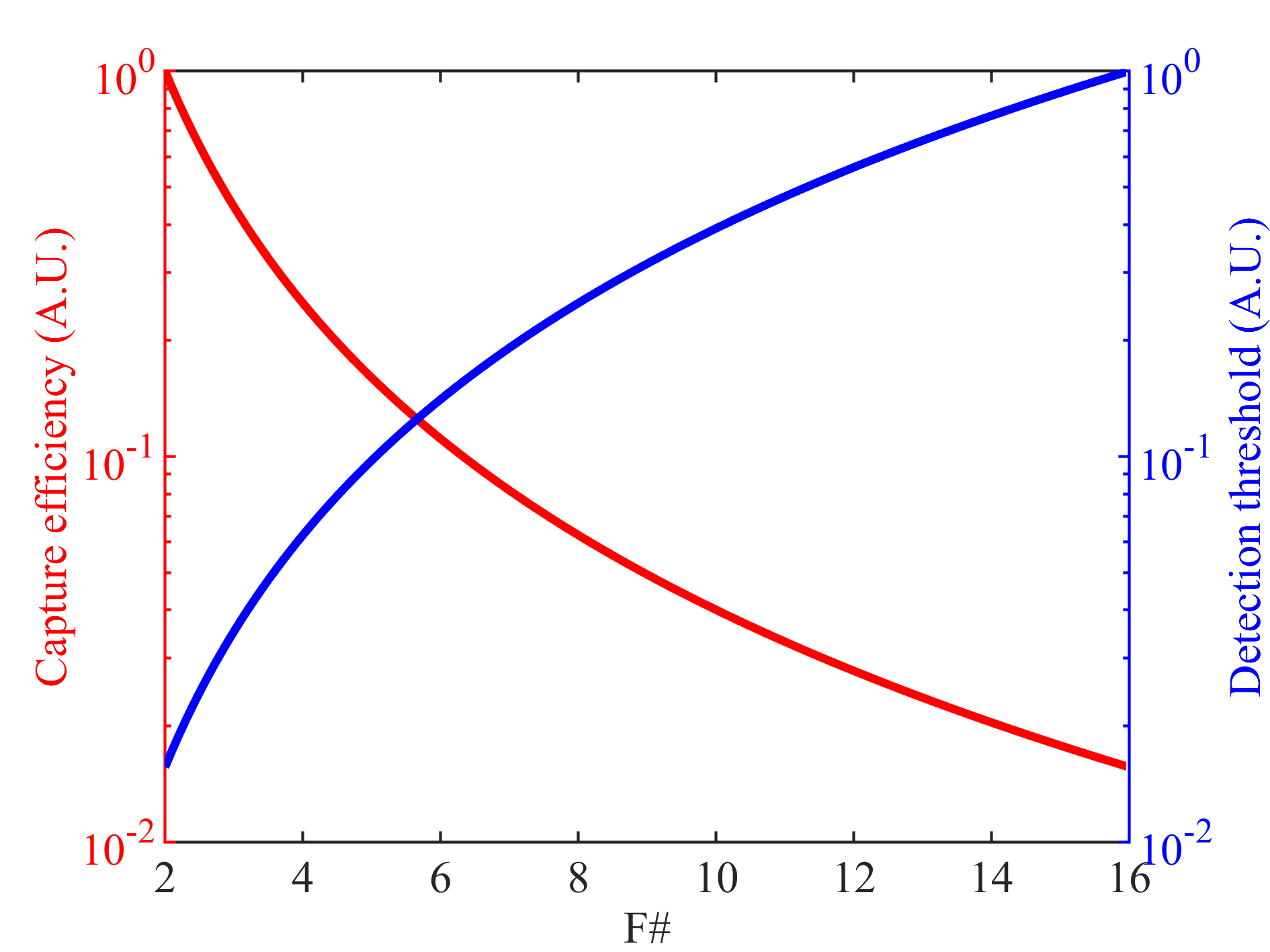}
		\caption{The photon collection efficiency and the detection threshold of the SFICS as a function of the F-number.}
		\label{fig:2}
	\end{figure}
	\linenumbers	

    \nolinenumbers	
\begin{figure*}[!htb]
	\centering
	\includegraphics[width=1\hsize]{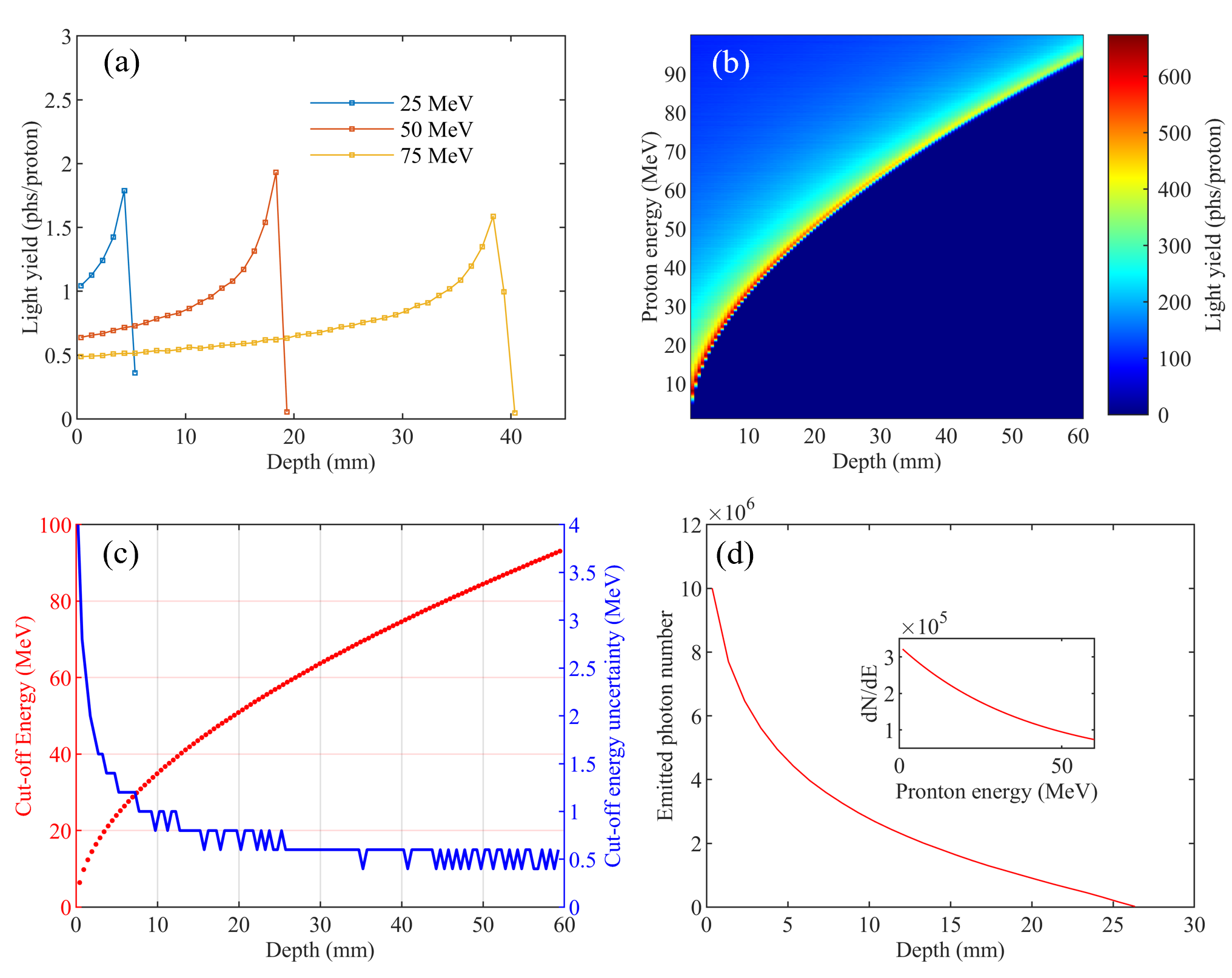}  
	\caption{(a) The emitted photons from scintillation fibers at depth of z for protons with incident energy of 25 MeV, 50 MeV, and 75 MeV, respectively. (b) Light yield response function $R(E,z)$ as a function of the energy of incident proton energy. (c) The cut-off energy and the uncertainty for scintillation fibers at different depths. (d) The variation of the emitted photon number with depth when SFICS is irradiated by a proton beam with an exponential decreasing energy spectrum. The inset figure depicts the energy spectrum of the proton beam.}
	\label{fig:3}
\end{figure*}
\linenumbers
	
	\subsection{Working principle}\label{subsec:working}
	
	As shown in Figure 1c, when the protons ballistically passes the scintillation fibers, they deposit energy and lead to the emission of scintillation photons. The photons are transported in scintillation fibers and eventually reach the edge face of the scintillation-fiber cube. The energy spectrum of incident proton beam can be written as $f(E,x,y)$. As the protons penetrate the scintillation-fiber cube and lose their energies, the energy spectrum at depth $z$ is changed and denoted as $g(E,x,y,z)$. Taking the XOZ plane as an example, neglecting the photon transportation losses in the scintillation fibers, the emitted photon number $N_\mathrm{SP}$ for a fiber $i$ (along the $x$ direction) is:
\begin{equation}
	\begin{aligned}
		N_\mathrm{SP}(y_i,z_i) &= \int_0^{E_\mathrm{cut}} \int_{-l/2}^{l/2} \iint_{\sigma_i} g(E,x,y,z) \\
		&\times \frac{dY_L}{dz} \, dx\,dy\,dz\,dE,
	\end{aligned}
	\label{eq:1}
\end{equation}
\textcolor{red}{	where $(y_i,z_i)$ is the coordinate of the center of the scintillation fiber, $l$ is the length of the scintillation fiber that ions pass by, $\sigma_i$ is the fiber cross-section, $E_\mathrm{cut}$ is the cut-off energy of the proton beam, and $\frac{dY_L}{dz}$ is the light yield per unit length, which is given by$^{47}$:}
	\begin{equation}
		\frac{dY_L}{dz} = \frac{S \cdot dE/dz}{1+kB \, dE/dz},
		\label{eq:2}
	\end{equation}
	where $S$ is the scintillation efficiency, and $kB$ is the quenching factor, $dE/dz$ is the proton stopping power.
	
	The emitted photons are recorded by the CCD and read out as CCD counts $Y_G(y_i,z_i)$:
	\begin{equation}
		Y_G(y_i,z_i) = \frac{K N_\mathrm{SP}(x_0,z_0)}{n(x_0,z_0)}.
		\label{eq:3}
	\end{equation}
	were $n(x_0,z_0)$ indicates the number of CCD pixels associated with the SF, and $K$ is the system response factor associated with the imaging system, which is inversely proportional to the solid angle of the SF with respect to the imaging lens.
	
	As described by Eq.~(\ref{eq:3}), the recorded CCD counts $Y_G$ are proportional to the scintillation photons number $Y_\mathrm{SP}$ and the system response factor $K$. Therefore, the detection range of the SFICS can be adjusted by adjusting the photon capture efficiency of the imaging system. One way is to adjust the optical aperture, i.e., the F-number of the imaging system.
	For a given F-number (F\#), the cone angle of the photons received by the lens is $\theta(F\#)$, the solid angle $\Omega(F\#)$ for photon reception by the lens is given by $\Omega(F\#)=2\pi(1-\cos(\theta(F\#)/2))$.
	
	As the detection threshold $DT(F\#)$ for scintillation photons is proportional to the receiving solid angle $\Omega(F\#)$ of the imaging system, $DT(F\#)$ of the SFICS exhibits an inverse relationship with $RE(F\#)$. As illustrated in Fig.~\ref{fig:2}, increasing the F-number results in a reduction in the relative photon collection efficiency due to the decreased lens aperture diameter, consequently elevating the detection threshold.

	\textcolor{red}{The diagnostic capability of the SFICS hinges on the dynamic range, defined as the ratio of the maximum non-saturating scintillation photon signal to the minimum detectable scintillation photon signal, which dictates the reliably measurable range of proton fluence.This is especially important for diagnosing proton beams from laser-plasma interactions, where the broad energy spectra and steeply decreasing proton fluence at higher energies lead to an exponential reduction in the corresponding scintillation photon intensity. } A wide dynamic range ensures accurate measurement of faint scintillation photon signals originating from high-energy protons, while preventing CCD saturation from intense scintillation photon signals generated by low-energy protons.
	
	Simply adjusting the lens aperture (e.g., increasing the F-number by reducing aperture size), while altering the overall scintillation light signal intensity, inherently shifts both the lower and upper detection thresholds in tandem. Consequently, the dynamic range remains unchanged because both detection thresholds scale proportionally. Similarly, the use of Neutral Density filters (ND filters), which are specifically designed to decrease the overall light intensity and avoid saturation, leads to a proportional attenuation of the scintillation light across all energy ranges. 
	
	To overcome these limitations, spatial segmentation filtering offers a practical approach. This technique involves dividing the scintillator cube surface into different regions and applying filters with varying Optical Densities (OD) to each region. Considering that lower-energy protons typically generate much stronger scintillation signals, higher-OD filters can be applied to areas where low-energy protons create intense signals, while lower-OD or transparent filters can be used for areas receiving high-energy protons with weaker signals. Although there is no strict quantitative relationship between proton energy and scintillation light output, this general correlation can be leveraged for filtering purposes.
    \nolinenumbers	
	\begin{figure}[!htb]
		\centering
		\includegraphics[width=1\hsize]{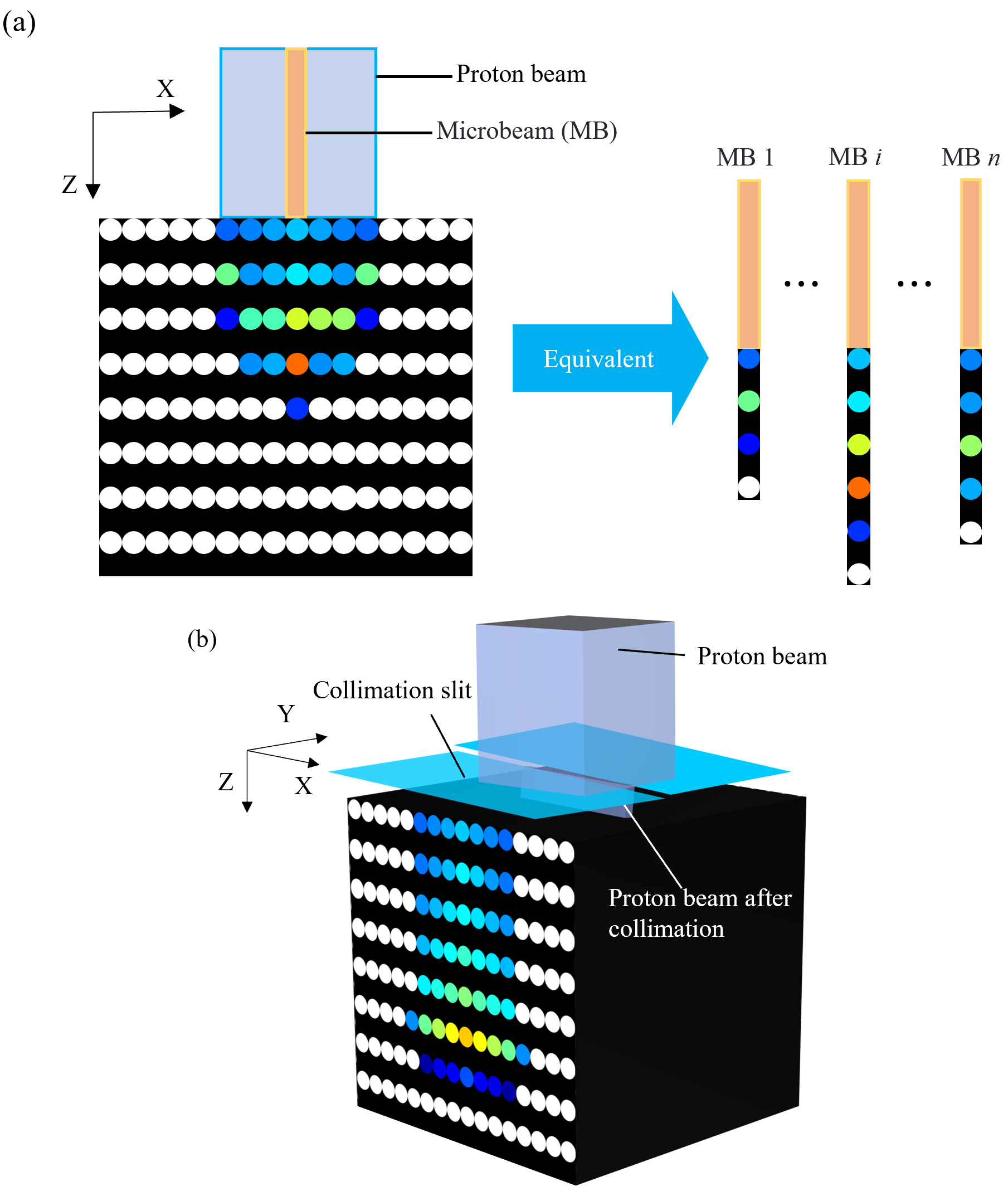} 
		\caption{(a) Schematic diagram of the energy spectrum retrieval method of the proton beam to be measured. (b) Collimation slit scheme.}
		\label{fig:4}
	\end{figure}
	\linenumbers	
	\subsection{Simulations of the output signals for uniform monoenergetic proton beams} \label{subsec:Simulations}
	
	Following the introduction of the working principle, this section focuses on simulating the detector response to monoenergetic, uniform proton beams—a foundational step for energy spectrum reconstruction. When protons pass through the scintillation-fiber cube, their energy deposition involves multiple physical effects: energy straggling, multiple Coulomb scattering, and non-linear light yield due to the quenching effect (dependent on the proton stopping power). These complexities render analytical inversion of the energy spectrum from the scintillation signal infeasible.
	
	To address this, we performed a Monte Carlo simulation using the Geant4 toolkit\cite{Agostinelli2003} encompassing the complete physical chain from proton transport through scintillation light generation and collection. The simulation framework initiates with a monoenergetic proton beam incident on the scintillation-fiber cube and systematically tracks the particle interactions throughout the detection process.
     \nolinenumbers	
\begin{figure*}[!t]
	\centering
	\includegraphics[width=1\hsize]{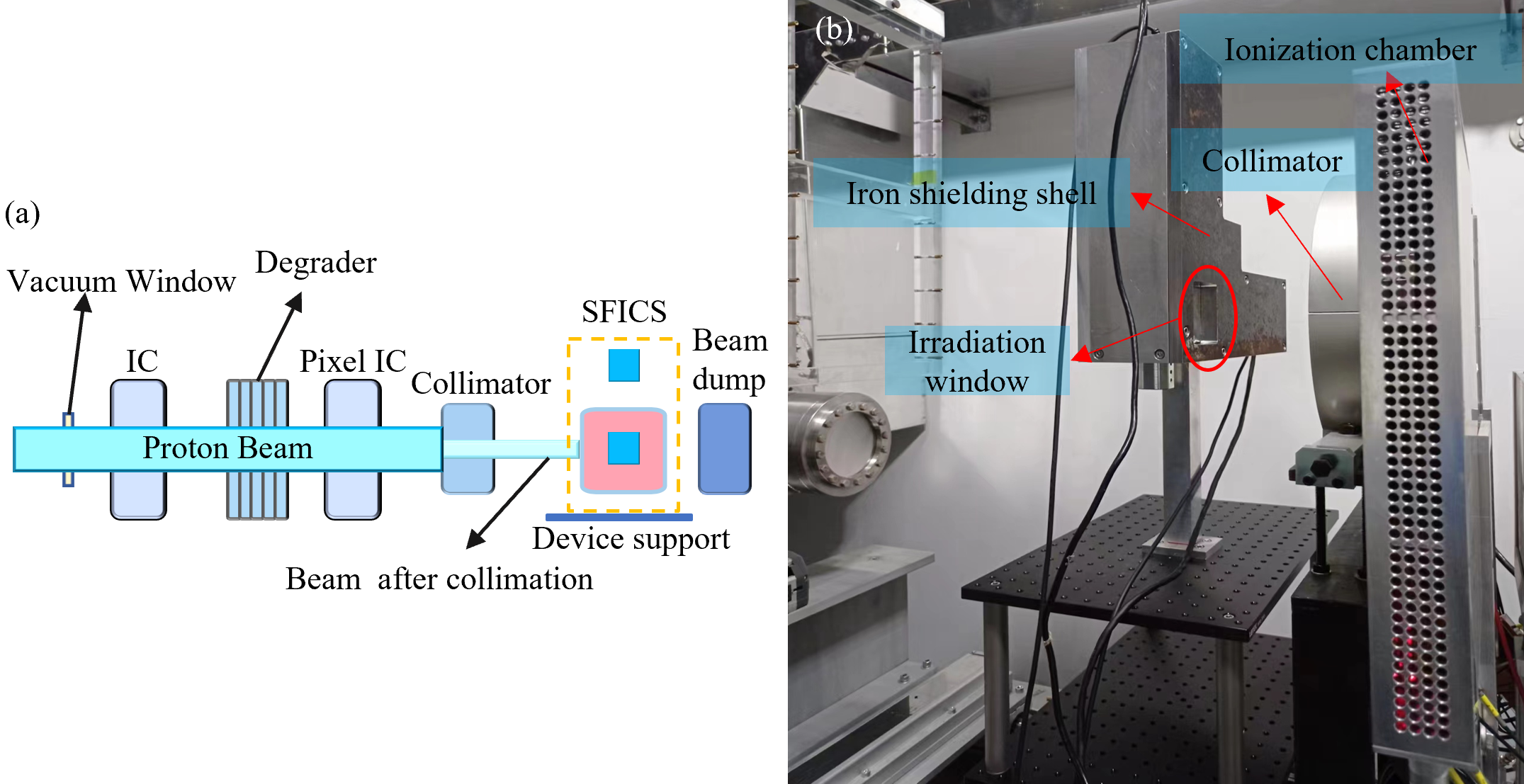}  %
	\caption{(a) Schematic of the calibration experiment setup. (b) Photograph of the SFICS placement during the experiment, positioned 13 cm downstream of the collimator.}
	\label{fig:5}
\end{figure*}
\linenumbers	
	
	The simulation incorporates detailed modeling of proton transport physics, primarily focusing on energy loss mechanisms (ionization and excitation), multiple Coulomb scattering, and range straggling. It then models the conversion of deposited energy into scintillation photons while incorporating ionization quenching effects, followed by detailed tracking of optical photon propagation through the scintillation fiber to its end.
	
	This detailed simulation generates a light yield response matrix $R(E,z)$, which quantifies the total detectable photons at depth $z$ per incident proton energy $E$. The matrix serves as the fundamental transfer function for subsequent spectrum retrieval algorithms.
	
	In the simulation, the scintillation-fiber cube, including a 100-$\upmu$m aluminum shielding film, was modeled based on the real structure of the SFICS. The quenching factor of the SF is set to 0.094 g${\cdot}$MeV$^{-1}{\cdot}$cm$^{-2}$ according to previous study\cite{Wang2012}. The predefined physics list QGSP\_BIC\_HP is utilized to model the interaction between protons and matter. Additional physical process model G4OpticalPhysics is employed to simulate the generation of scintillation photons.
	
	Fig.~\ref{fig:3}a shows the simulated light yield on the side face of the scintillation-fiber cube along the $z$ direction generated by three monoenergetic proton beams, which exhibit a Bragg peak shape. The higher the proton energy, the deeper the position of the peak. As shown in Figure~\ref{fig:3}b, the light yield response function $R(E,z)$ is presented for energies ranging from 1 MeV to 100 MeV, with a sampling interval of 0.1 MeV. It can serve as the data basis for the retrieve of the proton energy spectra. In the figure, it can be observed that the deepest scintillation fiber that emits light is determined by the energy of the incident proton, which can be used as an estimation of the cut-off energy of the incident proton beam.
	
	Considering a case where the coordinate of the deepest light-emitting scintillation-fiber layer is $z_n$, the corresponding cut-off energy is $E_\text{cut}(z_n)$. \textcolor{red}{The uncertainty of the cut-off energy is defined as $\Delta E_\text{cut}(z_n) = E_\text{cut}(z_{n+1}) - E_\text{cut}(z_n)$, representing the energy difference between adjacent layers. The relative energy uncertainty is $\eta(z_n) = \Delta E_\text{cut}(z_n) / E_\text{cut}(z_n)$.} Fig.~\ref{fig:3}c depicts the simulated cut-off energies for scintillation fibers locating at different depths. It can be seen that the measurable energy range of the SFICS is 6-93 MeV. With the increase of the proton energy, the uncertainty decreases. \textcolor{red}{For protons with energy higher than 60~MeV, the relative energy uncertainty is better than 1\%. In the 80~MeV region, the typical energy interval between adjacent layers is approximately 0.5~MeV, yielding $\eta \approx 0.6\%$.}
	
	We also simulate the photon number distribution of the SFICS irradiated by a proton beam with an exponential decreasing energy spectrum (see Fig.~\ref{fig:3}d). The results show that the emitted photons exponentially decrease with depth. Due to the significant number difference between the high-energy and low-energy protons, the shape of the curve of the emitted photon number as a function of depth is primarily determined by the energy spectrum.
   
	\nolinenumbers	
\begin{figure*}[!htb]
	\centering
	\includegraphics[width=1\hsize]{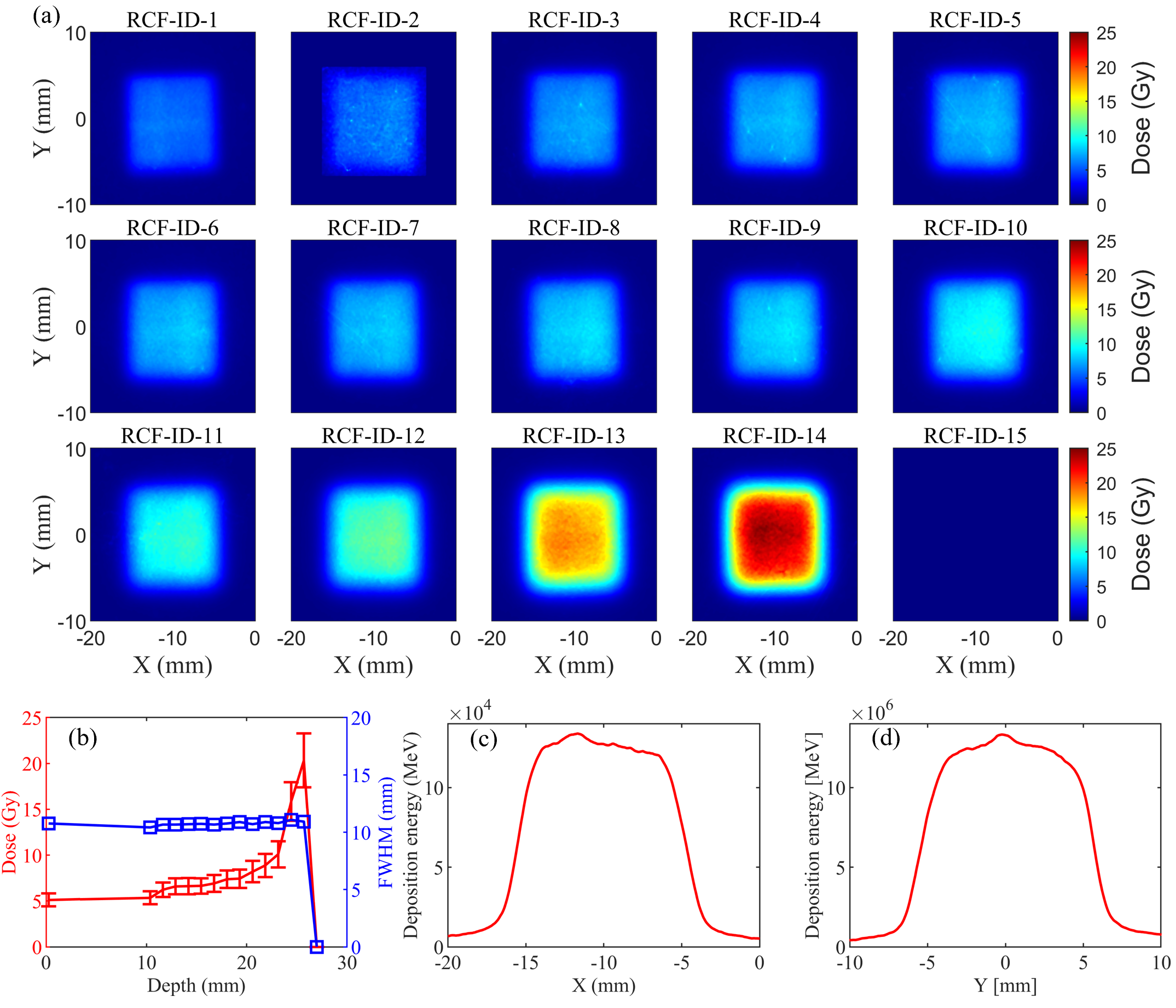}  
	\caption{(a) Two-dimensional dose distribution for each RCF. (b) Distribution curve of average dose and full width at half maximum of beam spot size with depth in the different RCF layers. (c) Energy deposition profile along the X direction obtained by integrating the energy deposition over the Y direction. (d) Energy deposition profile along the Y direction obtained by integrating the energy deposition over the X direction.}
	\label{fig:6}
\end{figure*}
\linenumbers	
\subsection{On the retrieve of the energy spectra} \label{subsec:retrieve}

The retrieve of the 2D differential energy spectrum $f(E,x,y)$ from SFICS measurements constitutes an inverse problem. The SFICS can only measure integrated radiation dose data along the scintillation fibers, and the observation angles are limited. The ill-posed nature of this problem makes it challenging to uniquely determine the full energy-spatial distribution, particularly for complex proton beams with correlated energy and position variations. However, in some scenarios, it remains feasible to retrieve the energy spectra. In this paper, we present some retrieval strategies as the first attempts.

\nolinenumbers	
\begin{figure*}[!htb]
	\centering
	\includegraphics[width=1\hsize]{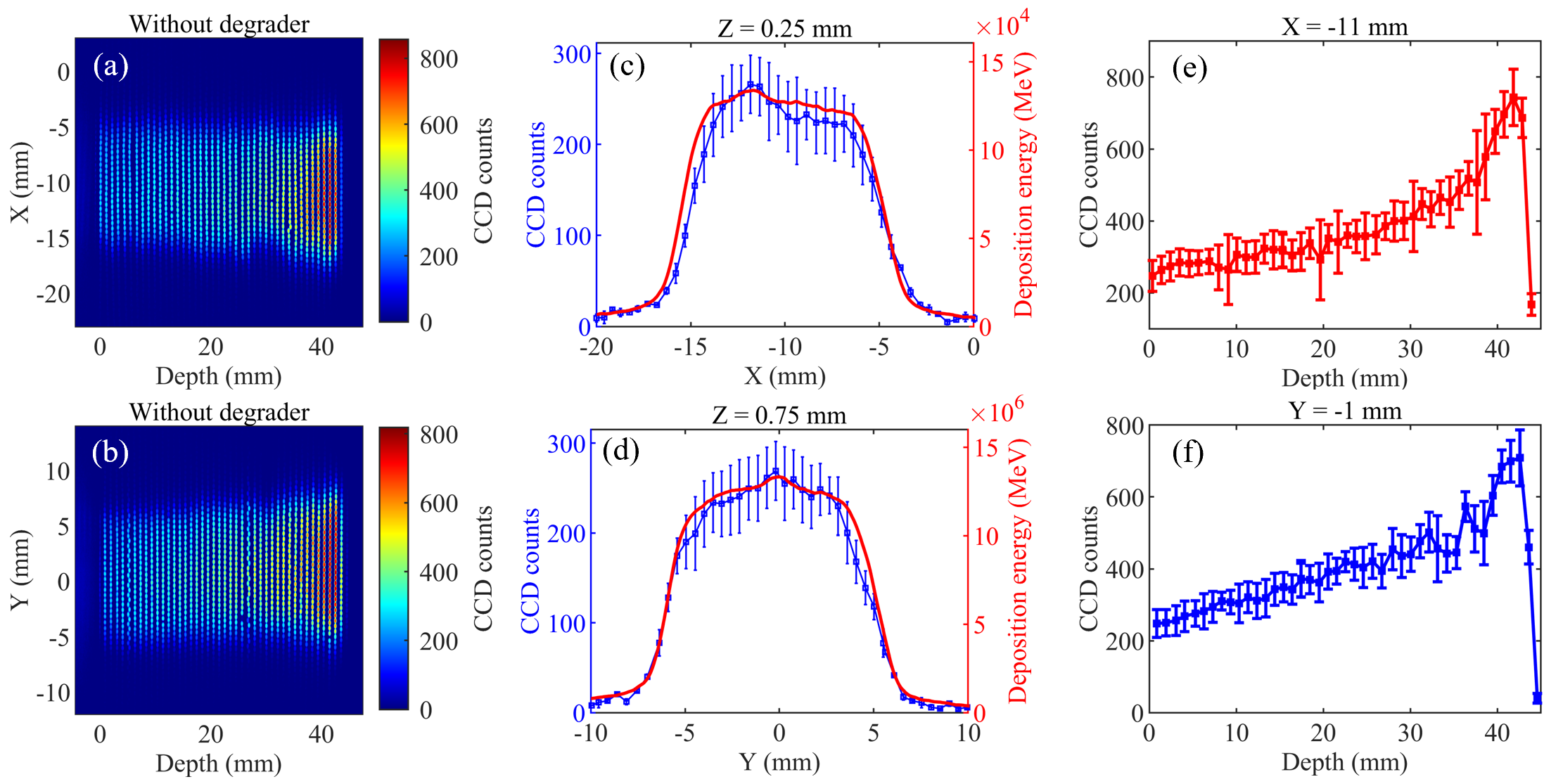}
	\caption{Spatial distribution of scintillation photons in the XOZ plane (a) and the YOZ plane (b). Comparison of 1D integrated proton beam profiles along the X-axis (c) and Y-axis (d) characterized by SFICS (blue lines) with those obtained from the RCF stack (red lines). Depth-dependent lineout profiles of scintillation light intensity at X=-11 mm in the XOZ plane (e) and at Y=-1 mm in the YOZ plane (f).}
	\label{fig:7}
\end{figure*}
\linenumbers
In the simplest case, if the spatial distribution of the proton beam is unvaried both in $x$ and $y$ direction, then the energy spectrum $f(E,x,y)$ simplifies to $f(E)$. One can use the SFICS as a simple scintillation stack spectrometer to retrieve the $f(E)$\cite{Nurnberg2009}. If the shape of energy spectrum is known in advance, the parameters of the spectrum can be obtained by using the least-squares method.

A harder case is that the spectrum of the proton beam along the $y$-axis is unvaried, and $f(E,x)$ follows a known slowly-varying distribution. This situation is equivalent to scintillation stack spectrometers along X axis (see Fig.~\ref{fig:4}a). This configuration effectively transforms the problem into an array of independent 1D inverse problems, where each $x$ position $x_k$ corresponds to a unique energy spectrum $f(E,x_k)$. The retrieval process then involves solving these parallel 1D inverse problems within a constrained least-squares optimization framework. The optimal spectrum parameters at each $x_k$ are determined by minimizing the following objective function:
\begin{equation}
	\label{eq:4}
	\begin{aligned}
		\text{MSE}&(x_k) = \frac{1}{M(x_k)} \sum_{j=1}^{M(x_k)} \Bigg( \int_0^{E_\text{cut}(x_k)} d \cdot Y_\text{FWHM} \\
		& \cdot f(E, x_k, \epsilon(x_k)) \cdot R(E, z_j)  \cdot /\ n_j \, dE - K\cdot\overline{Y_G^j} \Bigg)^2,
	\end{aligned}
\end{equation}
where $M(x_k)$ represents the number of scintillation fibers emitting light at position $x_k$, $E_\text{cut}(x_k)$ is the cut-off energy of the proton beam at position $x_k$, $d$ represents the diameter of the scintillation fibers, $Y_\text{FWHM}$ refers to the full width at half maximum (FWHM) of the one-dimensional transverse profile of the CCD count distribution in the YOZ plane, $R(E,z_j)$ represents the light yield response function of the SF at depth $z_j$, $\epsilon(x_k)$ is the parameter space to be retrieved for the proton energy spectrum $f(E,x_k,\epsilon(x_k))$ at position $x_k$, $n_j$ and $\overline{Y_G^j}$ represent the number of pixels and average CCD counts, respectively.

The Levenberg-Marquardt algorithm\cite{Murthy2015}, which has good convergence performance, can be used to solve this least-squares problem, with the iterative formula at the $(n+1)$th iteration given by:
\begin{equation}
	\epsilon_{n+1}(x_k) = \epsilon_n(x_k) - (H + \alpha I)^{-1} \nabla \text{MSE}(\epsilon_n(x_k)) \label{eq:5}
\end{equation}
where $H$ is the Hessian matrix of $\text{MSE}(x_k)(\epsilon_n(x_k))$, $I$ is the identity matrix, $\nabla$ denotes the gradient operator, and $\alpha$ is the damping factor. During the iteration, should the new parameter update lead to a reduction in the MSE, the damping factor is decreased; otherwise, it is increased.

The Levenberg-Marquardt algorithm provides an adaptive optimization framework that dynamically adjusts the damping factor $\alpha$ to balance gradient descent and Gauss-Newton iterations. This flexibility is particularly advantageous in handling noisy experimental data or spatially overlapping signals from scintillation fibers. Nevertheless, the effectiveness of the algorithm strongly depends on selecting an appropriate spectral model. Specifically, when applied to quasi-monochromatic beams, the Gaussian distribution assumption is well-suited to represent typical energy spreading phenomena, facilitating stable convergence. Conversely, attempts employing exponential forms generally exhibit deteriorating convergence behavior, ultimately failing to yield physically meaningful solutions due to inherent model-form incompatibilities.

If the spatial distribution of the proton beam in the $Y$ direction is varied as well, it’s highly challenging to simultaneously obtain the complete spatial-resolved spectra. Nevertheless, we can use a slit to measure the spectra in a slice spectrum $f(E,x,y_s)$ (see Fig.~\ref{fig:4}b), where $y_s$ denotes the position of the slit in the $Y$ direction. If the proton beam is stable for different bunches, the overall 2D differential energy spectrum $F(E,x,y)$ can be obtained by continuously moving the slit along the $y$ direction.

\section{Calibration of the SFICS} \label{sec:calibration}

In order to confirm the simulation results and obtain the response factor $K$ in Eq. (\ref{eq:3}) of the manufactured SFICS, we conducted the calibration experiments using proton beams from a synchrotron accelerator.
\nolinenumbers	
\begin{figure*}[!htb]
	\centering
	\includegraphics[width=1\hsize]{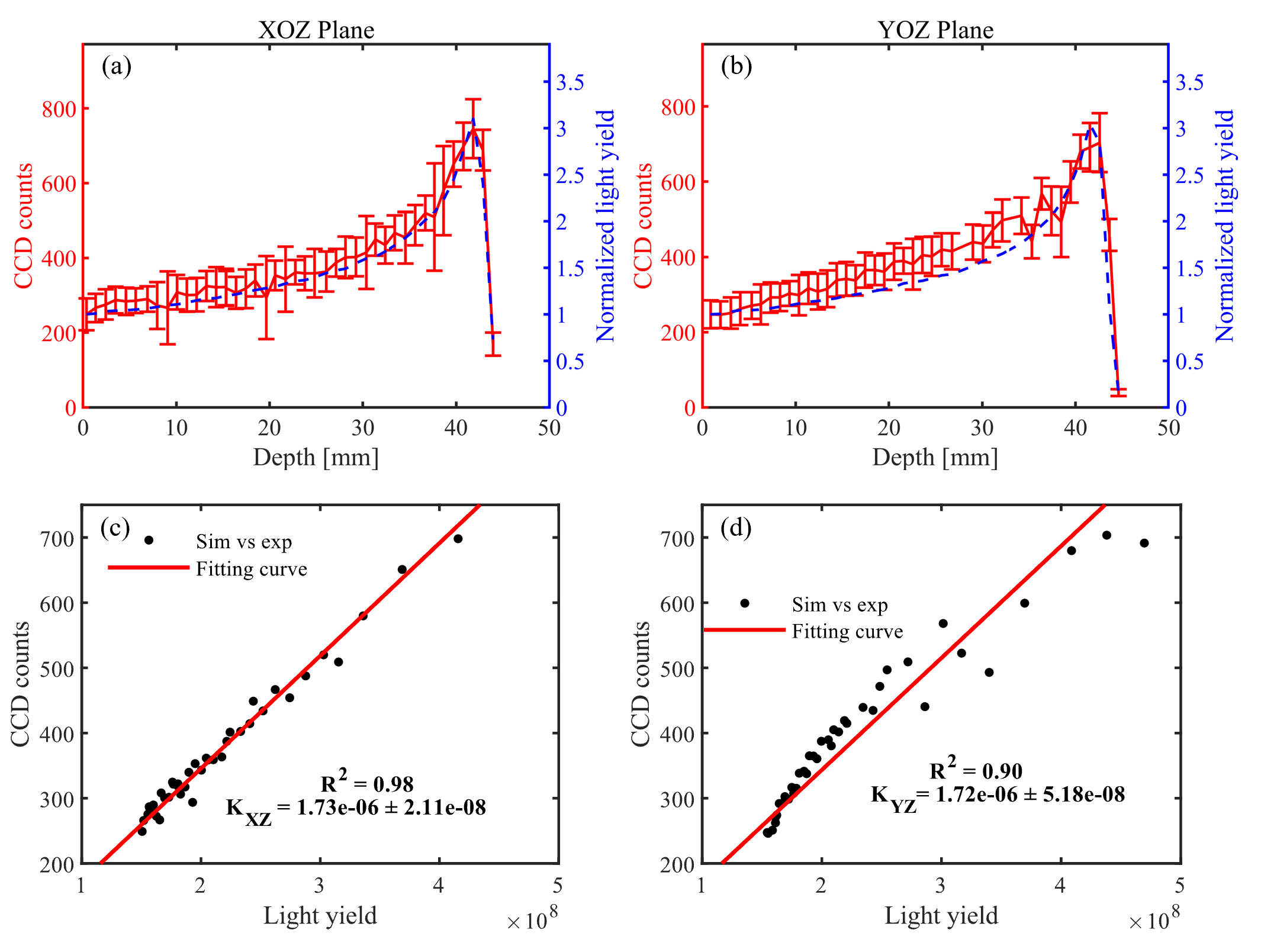}
	\caption{Depth-dependent lineout profiles of scintillation light intensity at X=-11 mm in the XOZ plane (a) and at Y=-1 mm in the YOZ plane (b) compared to the simulated light yield. Fit curves of CCD counts (exp) versus simulated (sim) light yield in the XOZ plane (c) and YOZ plane (d). The red line represents the fitted proportional curve. $K_{\text{XZ}}$ and $K_{\text{YZ}}$ correspond to the response factors of the SFICS in the XOZ and YOZ planes.}
	\label{fig:8}
\end{figure*}
\linenumbers	

\subsection{Proton beams used for calibration} \label{subsec:calib_beams}

The experiments were performed at the Xi'an 200 MeV Proton Application Facility (XiPAF)\cite{Wang2022}. \textcolor{red}{The energy of the proton beam at the experimental terminal (Fig.~\ref{fig:5}a) was set to 80 MeV, with a fluence of $\sim$2.0$\times$10$^9$ p/cm$^2$.} The proton beams sequentially passed (along the Z-direction) through the ionization chamber (intensity monitor), the degrader, the anode-partitioned pixel ionization chamber (profile monitor), and the collimator before they irradiate the SFICS. As shown in Fig.~\ref{fig:5}b, during the experiment, the SFICS was placed 13 cm behind the collimator. The side length of the square aperture was 1 cm. The CCD cameras of the SFICS was covered with a 20 mm thick iron shell to reduce radiation exposure to the CCD chip.

In the experiment, in addition to the ionization chamber, a RCF detector stack was used as well to accurately measure the energy spectrum of the proton beams. The RCF detector stack consisted of 16 RCF films and several aluminum plates of varying thickness, arranged in a specific order. \textcolor{red}{The RCF stack detector was placed at the irradiation window and irradiated by the proton beam, with a nominal energy of 80 MeV and a fluence of 6.43$\times$10$^9$ p/cm$^2$.}

The irradiated RCF films were scanned using an Epson scanner, resulting in 16 images in 800 dpi, 48-bit RGB uncompressed format. The dose deposited by the proton beam in each RCF film was calculated using the RCF dose calculation methods described in references\cite{Xu2019, Reinhardt2012, Gueli2015}.

The spatial distributions of the dose in different RCFs are shown in Figure~\ref{fig:6}a. The Bragg peak at the depth of $\sim$26 mm can be clearly seen. As the proton beam exits the collimator and reaches the RCF detector stack, it inevitably diffuses, resulting in a slightly larger beam spot size of 10.8 mm (as shown in Fig.~\ref{fig:6}b). \textcolor{red}{Consequently, the fluence at the surface of the RCF detector stack was lower than the fluence measured by the ion chamber.} To facilitate comparison with the SFICS measurement results, Figures~\ref{fig:6}c and~\ref{fig:6}d present the energy deposition profile of the proton beam along the Y and X axes, respectively, which indicate the spatial uniformity of the proton beam.
\nolinenumbers	
\begin{figure*}[!htb]
	\centering
	\includegraphics[width=1\hsize]{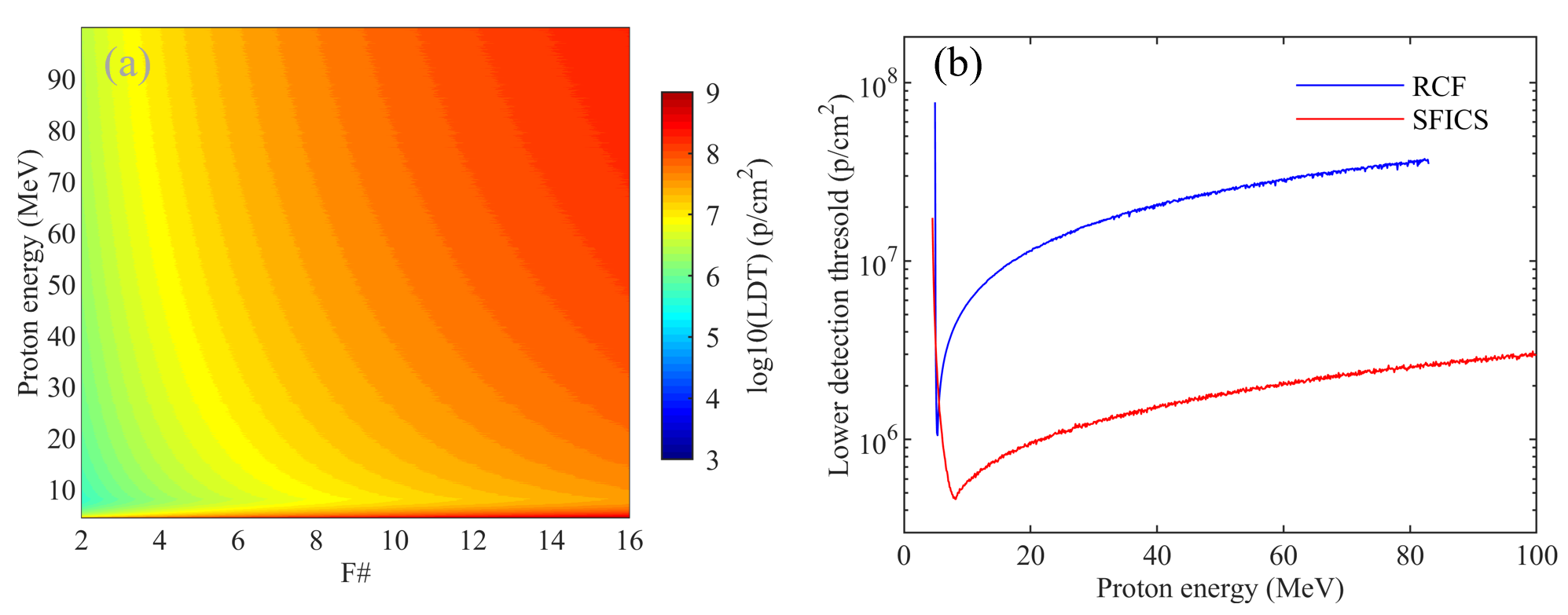}
	\caption{(a) The lower detection threshold of the SFICS for monoenergetic proton beams with varying energies at different F-numbers (b) A comparison of the detection threshold of the SFICS at F\#=2 with the RCF detector stack.}
	\label{fig:9}
\end{figure*}
\linenumbers	

The dose deposition of the proton beam in the RCFs, as illustrated in Fig.~\ref{fig:6}b, can be utilized to retrieve the proton beam energy spectrum. The problem of the retrieval method for RCF stacks can still be considered an ill-posed inverse problem, which can be solved by the regularized least-squares method. For the Gaussian energy distribution, the mean square error function MSE$_\text{RCF}$ is given by:

\begin{equation}
	\label{eq:6}
	\begin{aligned}
		\text{MSE}_\text{RCF} &= \sum_{i=1}^{N} \left( \int_0^{E_\text{cut}} f(E) \cdot R_i(E) \, dE - \Delta E_i \right)^2 \\
	\end{aligned}
\end{equation}

\begin{equation}
	\label{eq:7}
	f(E) = a_0 \exp\left[-\frac{(E-E_0)^2}{2\sigma^2}\right]
\end{equation}

where $f(E)$ represents the energy spectrum of the proton beam, $\Delta E_i$ is the total energy deposited by the proton beam in the $i$-th RCF, $a_0$ is a fit parameter affecting the number of protons, $E_0$ is the central energy of the spectrum, $\sigma$ is the standard deviation of the energy, and $R_i$ is the response curve representing the total energy deposited by a unit proton beam in the $i$-th RCF.

Using the L-M algorithm described above, we obtain the parameters in Eq. (\ref{eq:7}) as $a_0=5.81\times10^9$, $E_0=78.9$ MeV, and $\sigma=0.3$ MeV. The retrieved energy is slightly lower than 80 MeV, the assigned output energy of the protons. This is due to the energy loss in the vacuum window, the air, and the ionization chambers.
\subsection{Response Calibration}\label{subsec:responsecal}
	
The images of scintillation photons recorded by the two cameras in the SFICS were initially processed using standard methods including background subtraction, Gaussian filtering, and corner darkness correction\cite{Almurayshid2017}. Fig.~\ref{fig:7}a and \ref{fig:7}b present the processed images in the XOZ and YOZ planes of the scintillation-fiber cube, respectively, providing a visual representation of the spatial distribution of the detected scintillation light. The SFICS enables the determination of the absolute spatial position of the proton beam with respect to the center of the scintillation-fiber cube's surface, which is taken as the origin. In this particular measurement, the offsets of the beam center was 11 mm in the X direction and 1 mm in the Y direction, respectively.
\nolinenumbers	
\begin{figure*}[!htb]
	\centering
	\includegraphics[width=1\hsize]{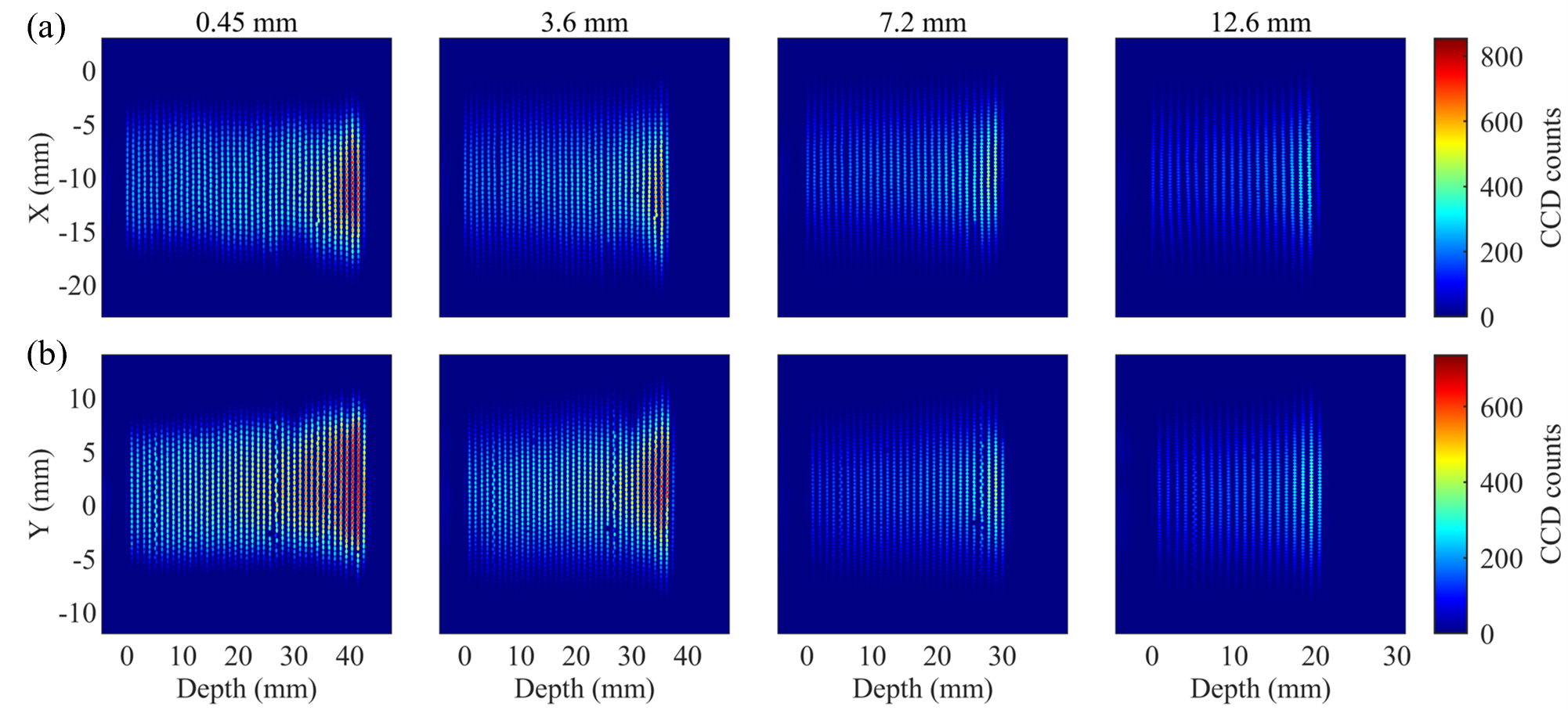}
	\caption{Spatial distribution of scintillation photons for degrader thicknesses of 0.45 mm, 3.6 mm, 7.2 mm, and 12.6 mm. (a) XOZ plane. (b) YOZ plane.}
	\label{fig:10}
\end{figure*}
\linenumbers	
\nolinenumbers	
\begin{figure*}[!htb]
	\centering
	\includegraphics[width=1\hsize]{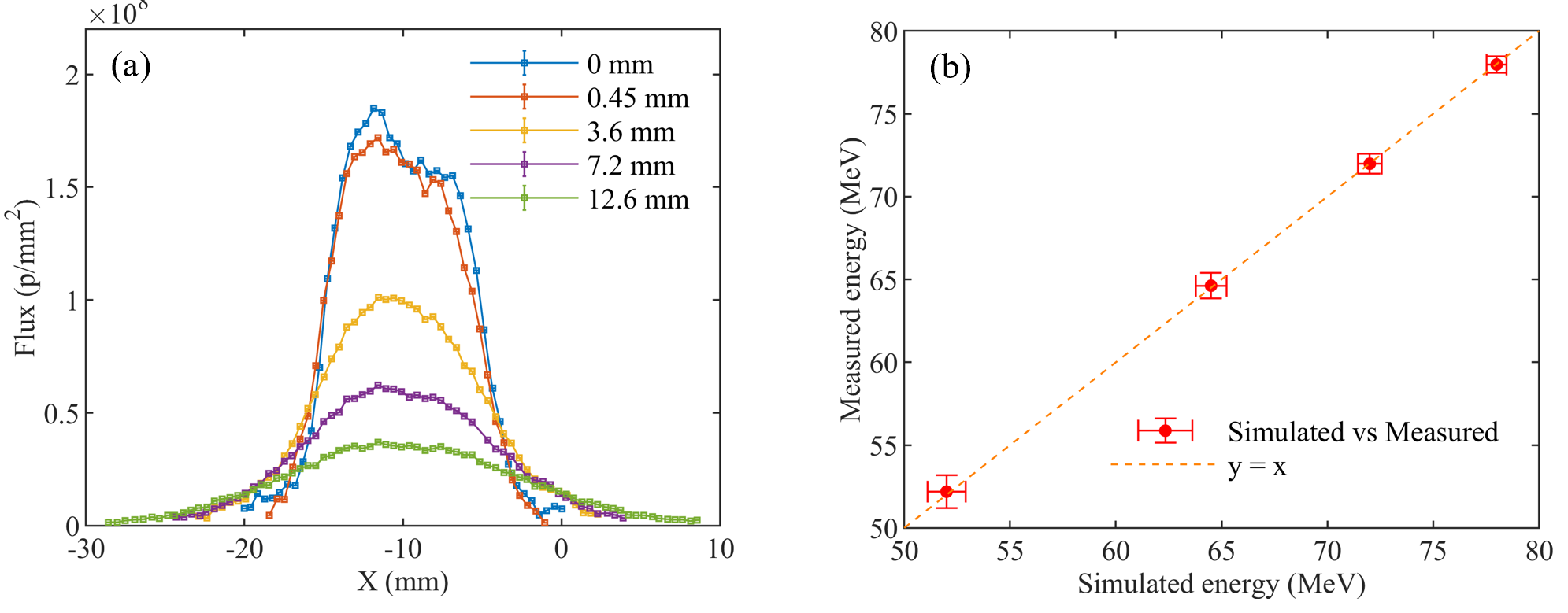}
	\caption{\textcolor{red}{(a) Comparison of proton fluence distributions along the X-direction for different degrader thicknesses.} (b) Comparison of the simulated central energy and energy spread of the proton beam after degradation with the values measured by the SFICS.}
	\label{fig:11}
\end{figure*}
\linenumbers	
The one-dimensional integrated proton beam profiles along the X and Y axes were derived from line scans of the scintillation light distribution at z=0.25 mm and z=0.75 mm (blue curves in Figs.~\ref{fig:7}c and \ref{fig:7}d). At these depths near the entrance surface of the SFICS, the protons had not yet undergone significant angle straggling. The beam profiles measured by the SFICS showed reasonable agreement with those obtained from the RCF stack (red curves in Figs.~\ref{fig:7}c and \ref{fig:7}d). The consistency between these two independent methods validates the capability of the SFICS in characterizing the spatial distribution of proton beams.

To determine the system response factor $K$ in Eq.~(3), a comparison between the CCD counts and the simulated light yield ($Y_p$) is necessary. The CCD counts along the depth direction at X = -11 mm and Y = -1 mm have been extracted and are presented in Fig.~\ref{fig:7}e and \ref{fig:7}f, which exhibit a Bragg peak shape. \textcolor{red}{To enhance the reliability of the simulation, the actual detected parameters, including fluence and energy, were input into Geant4 for computation. The proton beam fluence at X = -11 mm and Y = -1 mm is estimated based on the 1D integral beam profiles measured by the RCF stack, along with the fluence detected by the ionization chamber, yielding results of 1.61$\times$10$^8$ p/cm$^2$ and 1.64$\times$10$^8$ p/cm$^2$, respectively.}

By comparing the experimental data with the normalized simulation data, we can obtain there is a strong linear correlation between the CCD counts and the normalized simulated light yield (see Fig.~\ref{fig:8}a and \ref{fig:8}b). By fitting the CCD counts to the simulated light yield (see Fig.~\ref{fig:8}c and \ref{fig:8}d), the system response factors $K$ of the SFICS in the XOZ and YOZ planes were determined to be 1.73$\times$10$^{-6}$ and 1.72$\times$10$^{-6}$ count/photons respectively. It is worth noting that the F-numbers of the two imaging systems were set to 2 in this experiment.

\subsection{Detection threshold}\label{subsec:detthreshold}

\textcolor{red}{Once the system response factor K has been calibrated, the lower detection threshold (LDT) for proton fluence can be quantified, elucidating the intrinsic sensitivity boundaries of the spectrometer.} With ambient light fully shielded, the system noise is dominated by the SFICS's camera readout noise (5 e$^-$ rms) and full-well capacity (10,000 e$^-$). Proton signals are deemed detectable when the signal-to-noise ratio (SNR) reaches 3.

\textcolor{red}{This proton fluence detection threshold depends on beam spot size. For this calculation, a proton beam size of 10 mm is considered.} The LDT calculated using the light yield response matrix of the scintillation-fiber cube is shown in Figure~\ref{fig:9}a. When F\#=2, the LDT is on the order of 10$^6$ p/cm$^2$. While at the F-number of 16, the LDT is on the order of 10$^8$ p/cm$^2$. As shown in Figure~\ref{fig:9}b, at F\#=2, it can detect proton beams as low as 4.5 MeV. Compared to the RCF detector stack used in the experiment (minimum detectable dose: 0.2 Gy), SFICS is one-order-of-magnitude more sensitivity for proton beams with energies higher than 8 MeV.

\section{Application of SFICS}\label{sec:application}
After calibration, the SFICS can be utilized to measure the energy spectra of proton beams using the retrieval method. Two application scenarios are presented below: the first is to measure the energy of a quasi-monoenergetic proton beam, and the second is to measure the proton beam with a broad energy distribution and spatially varied energy spectrum.

\subsection{Energy measurement of uniform quasi-monoenergetic proton beam}\label{subsec:uniformbeam}
This experiment employed energy degraders made of Al plates with varying thicknesses to generate quasi-monoenergetic proton beams of different energies. Fig.~\ref{fig:10} shows the images recorded by the SFICS in the XOZ and YOZ plane, respectively. It can be seen that the maximum depth with scintillation emission decrease with increasing degrader thickness (Fig.~\ref{fig:10}a and \ref{fig:10}b).\textcolor{red}{ Additionally, due to the interaction with the scintillation-fiber cube, the proton beam exhibits both an expansion in spot size and a reduction in proton fluence with increased Z.} These observed phenomena directly result from energy loss and multiple Coulomb scattering, consistent with the physical principles modeled in the simulation.

\textcolor{red}{The previously described energy spectrum retrieval method was employed to extract CCD count profiles along the depth for various Y positions. Inputting these profiles into the retrieval algorithms yielded the energies and fluences of the quasi-monoenergetic proton beams.} A key assumption in this process is that $f(E,x_k)$ follows a Gaussian distribution due to the quasi-monoenergetic nature of the beam after energy degradation. This assumed Gaussian energy spread is independent of the proton beam's initial spatial profile.
 \nolinenumbers	
\begin{figure*}[!htb]
	\centering
	\includegraphics[width=1\hsize]{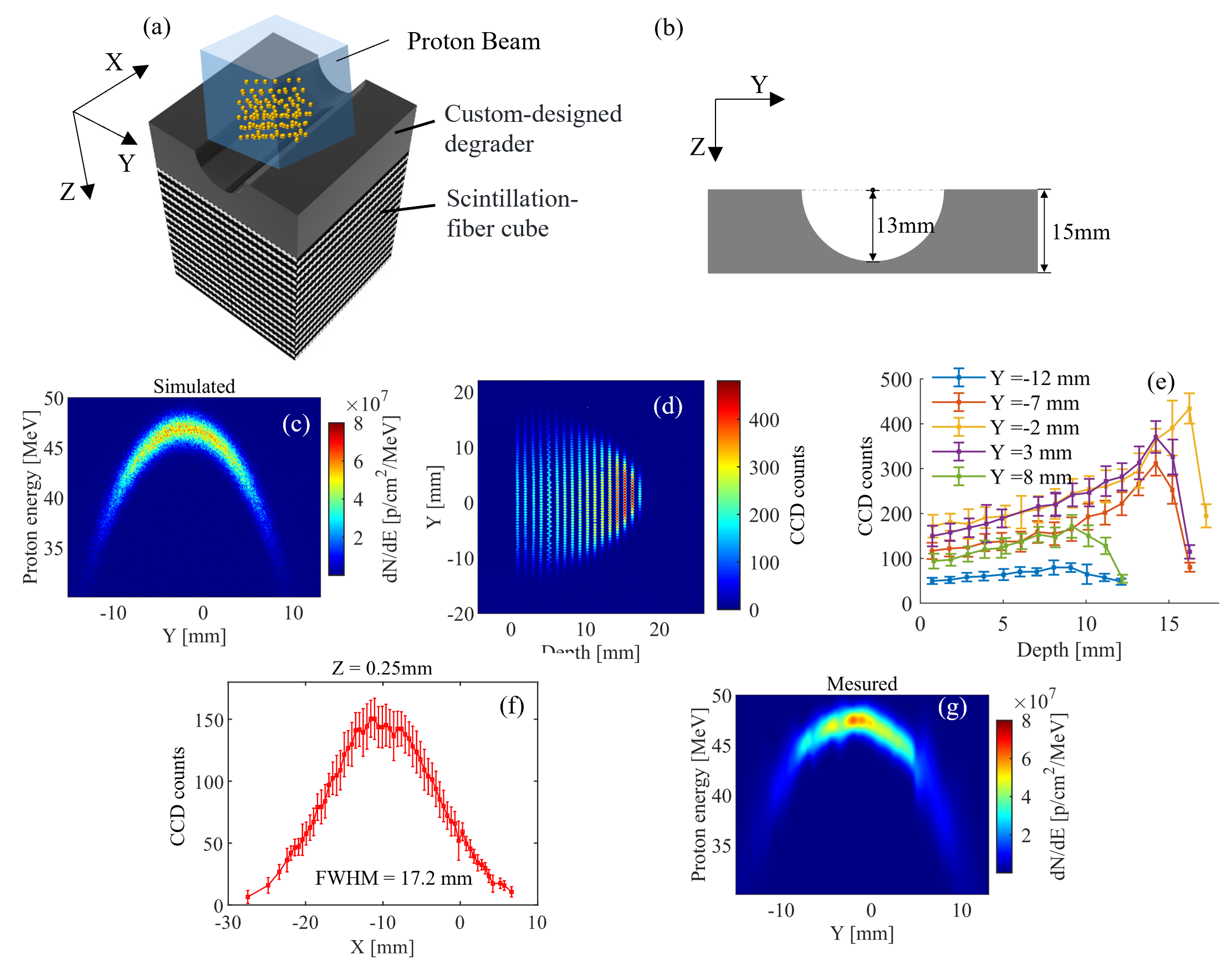}
	\caption{(a) Schematic of the proton beam irradiating the SFICS after being modulated by a custom-designed degrader. (b)Cross-sectional view of the custom-designed degrader. (c) Simulated energy spectrum distribution of the proton beam in the Y direction at the scintillation-fiber cube. (d) Symmetrical peak distribution of scintillation photons in the YOZ plane. (e) CCD count distributions along depth direction extracted at different Y positions. (f) 1D transverse profile of the proton beam in the X direction characterized by scintillation light intensity. (g) Reconstructed mean 2D energy spectrum distribution of the proton beam in the Y direction at the scintillation-fiber cube. }
	\label{fig:12}
\end{figure*}
\linenumbers	

\textcolor{red}{As a result, the proton fluence distribution along the X-axis near the surface of the scintillation-fiber cube is calculated, as shown in Fig.~\ref{fig:11}a. With a minimal degrader thickness of 0.45 mm, the proton beam exhibits only slight alterations in size and fluence upon reaching the scintillation-fiber cube. As the degrader's thickness increases, the initial beam profile, which deviates from a Gaussian distribution, progressively evolves towards a Gaussian shape. As the profile transforms, the fluence decreases.} The observed evolution of the beam profile is consistent with beam broadening due to multiple scattering. Furthermore, as the degrader thickness increases from 0.45 mm to 12.6 mm, the Full Width at Half Maximum (FWHM) of the beam spot increases to approximately 11 mm, 12.5 mm, 14.5 mm, and 16.5 mm, respectively.

\textcolor{red}{In addition to proton fluence, the energy spectrum inversion method also enabled the determination of the proton beam's central energy and energy distribution.} The retrieval accuracy of the SFICS is enhanced at higher proton beam energies. As shown in Fig.~\ref{fig:11}b, the retrieved values demonstrate good agreement with simulated values. For instance, at a simulated central energy of 78.0 MeV, the reconstructed value closely aligns with the simulation result. High accuracy is maintained even at lower energies; at a simulated energy of 52 MeV, the measured value deviates from the simulated value by only 0.2 MeV.

\subsection{Measurement of non-uniform proton beam}\label{subsec:nonuniformbeam}
To validate the spectrum retrieval method, we used a custom-designed energy degrader to obtain a spatially non-uniform and approximately parallel proton beam with a broad energy distribution (see Fig.~\ref{fig:12}a). As shown in Fig.~\ref{fig:12}b, the cross-sectional view of the custom-designed degrader is depicted. It is fabricated by milling a 13-mm-radius semi-cylindrical groove from a 15-mm-thick aluminum block. The custom-designed degrader is designed to enable the proton beam to undergo varying levels of energy degradation at different positions along the Y-direction.

The proton beam first traversed the 12.6-mm-thick degrader, then passed through the custom-designed degrader, and finally irradiated the scintillation-fiber cube. \textcolor{red}{The experimental setup was modeled in Geant4, with an initial beam fluence of 2.08$\times$10$^8$ p/cm$^2$ (measured by the ionization chamber).} The simulated energy distribution of the proton beam versus the spatial Y-axis at the surface of the scintillation-fiber cube is shown in Fig.~\ref{fig:12}c. This indicates that proton beams can be subdivided into quasi-monoenergetic proton microbeams at different positions.

In the experiment, the image recorded by the SFICS in the YOZ plane are shown in Fig.~\ref{fig:12}d. The image displays a symmetrical peak centered at Y = 2 mm, originating from the non-uniform energy degradation of the proton beam along the Y-direction.

At different positions in the Y direction, the proton beam passed through different thickness within the custom-designed degrader, leading to differential energy degradation. According to the energy spectrum retrieval method shown in Fig.~4a, the CCD count distribution along depth direction at different positions in the Y direction was extracted. Fig.~\ref{fig:12}e shows the CCD count distributions at five different Y positions, all of which exhibit Bragg peak shapes. Therefore, the energy spectra at different positions along the Y direction still approximately follow a Gaussian energy distribution. By substituting the CCD counts at different positions in the Y direction and the calibrated response factor $K$ into Eq.~(\ref{eq:4}), the corresponding mean square error function was obtained, and the corresponding energy spectra were retrieved iteratively using the method described in Eq.~(\ref{eq:5}).

Based on the 1D integral beam profile in the XOZ plane (see Fig.~\ref{fig:12}f), the FWHM of the beam spot in the X direction was determined to be 17.2 mm. Integrating $F(E,y_k)$ together, the retrieved energy spectrum $F(E,y)$ is obtained as shown in Fig.~\ref{fig:12}g. Overall, the retrieved result aligns well with the simulated energy spectrum, validating the effectiveness of the energy spectrum retrieval method and confirming that the SFICS can retrieve the 2D differential energy spectrum of spatially non-uniform beams under some specific conditions.
 
	\section{Conclusion}\label{sec:Conclusion}
	In summary, we have developed and validated a novel Scintillation-Fiber-Cube Spectrometer (SFICS) designed to simultaneously provide high spatial and energy resolution for proton beam diagnostics. \textcolor{red}{The SFICS integrates orthogonally layered scintillation fibers with high-resolution imaging systems to achieve quasi-2D spatial measurements with a pixel size of 0.5~mm for beam profile reconstruction and relative energy uncertainty better than 1\% for protons above 60~MeV.} Calibration experiments using synchrotron-accelerated proton beams and Geant4 simulations demonstrate the SFICS's capability to characterize complex energy spectra and varying spatial profiles. Compared with RCF stack detectors, the SFICS achieves real-time online proton beam diagnostics while maintaining comparable energy resolution and detection sensitivity, establishing itself as a promising alternative to RCF-based spectral diagnostics systems.
	
	Future development by using segment filters can be performed to enlarge the SFICS's dynamic range to accommodate the broad energy spectra and pronounced spectral gradients of laser-accelerated proton beams Advanced computation methods, particularly machine learning algorithms, may be utilized to enhance the efficiency and precision of the retrieved energy spectra.

\section*{Acknowledgments}

\textcolor{red}{This work was supported by the National Grand Instrument Project (No.~2019YFF01014402), the National Key Research and Development Program of China (No.~2024YFF0726304), the National Natural Science Foundation of China (No.~12575257, 12205008), Beijing Natural Science Foundation (No.~1252019), and the National Key R\&D Program from the Ministry of Science and Technology of China (No.~2019YFE0114300). W.~Ma acknowledges support from the National Science Fund for Distinguished Young Scholars (No.~12225501). The Monte Carlo (GEANT4) simulations were carried out on the High-Performance Computing Platform of Peking University.}

\end{document}